\documentclass[11pt,a4paper,notitlepage]{article}

\usepackage{epsfig}

\begin{document}

\title{Instanton Dominance of Topological Charge Fluctuations in QCD?}

\renewcommand{\thefootnote}{\fnsymbol{footnote}} \author{I.
  Hip$^a$\footnote{Corresponding author. \textsl{E-mail address:}
    hip@theorie.physik.uni-wuppertal.de},
  Th. Lippert$^a$, H. Neff$^b$, K. Schilling$^a$ and W. Schroers$^a$\\
  \textit{$^a$Fachbereich Physik, Bergische Universit\"at, Gesamthochschule
    Wuppertal} \\
  \textit{Gau\ss stra\ss e 20, 42097 Wuppertal, Germany} \\
  \textit{$^b$ NIC, c/o Research Center J{\"u}lich, D-52425 J{\"u}lich} \\
  \textit{and DESY, D-22603 Hamburg, Germany}}
  \date{2. May 2001}
\maketitle \renewcommand{\thefootnote}{\arabic{footnote}}

\begin{abstract}
  We consider the local chirality of near-zero eigenvectors from Wilson-Dirac
  and clover improved Wilson-Dirac lattice operators 
  as proposed recently by Horv\'ath et al.  We studied
  finer lattices and repaired for the loss of orthogonality due to the
  non-normality of the Wilson-Dirac matrix. As a result we do see a clear
  double peak structure on lattices with resolutions higher than 0.1 fm.  We
  found that the lattice artifacts can be considerably reduced by exploiting
  the biorthogonal system of left and right eigenvectors.  We conclude that
  the dominance of instantons on topological charge fluctuations is not
  ruled out by local chirality measurements.
\end{abstract}

\section{Introduction}
\label{sec:introduction}
It is generally accepted that instanton field configurations play an important
part in hadron phenomenology~\cite{Shuryak:1982ff}. The space-time structure
of instantons has been the target of a variety of lattice investigations in
the past~\cite{Negele:1999ev}. Recently Horv\'ath et al.~\cite{Horvath:2001ir}
have added a very interesting contribution to this discussion about the r\^ole
of instantons.  They studied the local chiral orientation of near-zero
eigenfunctions on an ensemble of (rather coarse grained) quenched QCD vacuum
configurations, but saw no evidence for instanton driven topological charge
fluctuations. On the other hand, by applying the techniques of
Horv\'ath et al.~to
overlap \cite{Neuberger:1998fp} instead of standard Wilson fer\-mions,
DeGrand and Hasenfratz arrived at the opposite
conclusion~\cite{DeGrand:2001pj}.

Wilson-type actions have the merit of being ultralocal which is desirable for
the study of local properties but at the cost of chiral symmetry breaking.  On
the other hand the overlap formulation has a remnant of chiral symmetry
\cite{Luscher:1998pq} but might wipe out local structures, as it implies
couplings beyond nearest neighbours \cite{Horvath:1998cm,Hernandez:1999et}.

In view of the implications of the above findings for our understanding of the
QCD vacuum structure we shall focus in this letter on possible lattice
artifacts due to the use of Wilson-type operators.  At finite
lattice spacing $a$ these
operators are non-normal.  In Section \ref{sec:local-chirality} we shall
remind the reader of some peculiarities encountered in the eigenmode expansion
due to this non-normality \cite{Smit:1987fn,Ivanenko:1998nb}.  The latter
induces a biorthogonal system of distinct right and left eigenvectors.  This
suggests to modify the local chiral orientation parameter $X$ proposed in
Ref.~\cite{Horvath:2001ir}.  The resulting lattice improved form,
$X_{\mbox{\tiny imp}}$,
is based on projecting the chiral reduction of the right
eigenvectors onto the {\it related left} eigenvectors.  In Section
\ref{sec:results-qed2} we shall demonstrate this improvement for the case of
QED2.  In Section \ref{sec:results-qcd} we shall apply $X_{\mbox{\tiny imp}}$
to the case of QCD.

\section{Local Chirality in Non-normal Scenario}
\label{sec:local-chirality}
Two commonly used discretizations of the Dirac operator are the
Wilson-Dirac matrix, $D_W$, and the  clover improved Wilson-Dirac
matrix, $D_{CW}$. Both matrices are non-normal
\begin{equation}
  [D,D^{\dagger}] \neq 0\; . \label{eq: normality}
\end{equation}
This fact implies that both matrices  are not diagonalizable by unitary
transformations. Yet their diagonalization can be achieved  by similarity
transformations with non-unitary matrices
\begin{equation}
  \Lambda = S^{-1} D S \; , \label{eq-sim}
\end{equation}
with $\Lambda$ being the diagonal matrix.  The columns of $S$ are then the
\textit{right} eigenvectors, $|R_i\rangle$, and the rows of $S^{-1}$ the
\textit{left} eigenvectors, $\langle L_i|$, of $D$ (see
e.g.~\cite{Golub:1996bo}),
\begin{eqnarray}
  D | R_i \rangle &=& \lambda_i | R_i \rangle, \nonumber \\
  \langle L_i | D &=& \lambda_i \langle L_i |\; .
\end{eqnarray}
It follows directly from $S^{-1}S=I$ that the left and right
eigenvectors are \textit{biorthogonal}
\begin{equation}
  \langle L_i | R_j \rangle = \delta_{ij}\; . \label{eq-biorth}
\end{equation}
In contrast, the right (or left) eigenvectors by themselves  \textit{do not}
form an orthogonal system.

This biorthogonality has to be taken into account when 
expressing  an abstract operator $A$ in terms of a
matrix representation by use of eigenfunctions
\begin{equation}
A = \sum_{i,j}  | R_i \rangle \langle L_i |
A   | R_j \rangle \langle L_j |\; . \label{eq:matrix-el}
\end{equation}
In the following we will consider the operator $( 1 \pm \gamma_5
)$, which can be expanded accordingly
\begin{equation}
1 \pm \gamma_5 = \sum_{i,j} | R_i \rangle \langle L_i | 1 \pm
 \gamma_5 | R_j \rangle \langle L_j | \label{eq:ch-projector}\; .
\end{equation}
(Actually, we will need only the diagonal elements $\langle
 L_i | 1 \pm \gamma_5 | R_i \rangle$).
Only in the continuum limit, where $D$ becomes
normal, the operators in Eqs.~(\ref{eq:matrix-el}) and (\ref{eq:ch-projector})
can be expanded in terms of right eigenvectors only.

Fortunately, though, one encounters no duplication of the costs to compute the
left eigenvectors of $D_W$ and $D_{CW}$ because these operators obey
the $\gamma_5$-hermiticity property
\begin{equation}
  D^{\dagger} = \gamma_5 D \gamma_5 \; . \label{eq:g5herm}
\end{equation}
This symmetry allows for a straightforward
construction of the left eigenvectors from  the right ones:
\begin{equation}
  \langle R_i | \gamma_5 D =
  \lambda^{\ast}_i \langle R_i | \gamma_5 \; .
\end{equation}
Hence the left eigenvector to  the eigenvalue $\lambda^{\ast}_i$ is 
nothing but the
hermitian conjugate of the right eigenvector to $\lambda_i$ multiplied
by $\gamma_5$ from the right:
\begin{equation}
  \langle L_i | D =  \langle R_i | \gamma_5 D =
  \lambda^{\ast}_i \langle L_i | \; .
\end{equation}

\paragraph{Horv\'ath et al.~method.}
In order to quantify the r\^{o}le of instantons in the topological charge
fluctuations in QCD Horv\'ath et al.~define the local ``chiral orientation''
parameter $X$ of an eigenmode $|\psi\rangle$ at every lattice point
(cf.~Eq.~(19) in \cite{Horvath:2001ir}),
\begin{equation}
  \tan \left(\frac{\pi}{4}(1 + X) \right) = \frac{|\psi_-|}{|\psi_+|}\; .
  \label{eq-X}
\end{equation}
The absolute values 
 $|\psi_-|$ and $|\psi_+|$ stand for~\footnote{We use $\psi_-$ and
  $\psi_+$ instead of their $\psi_L$ and $\psi_R$ to avoid confusion with
  the left and right eigenvectors.}
\begin{equation}
|\psi_{\pm}|^2 \equiv \langle  \psi_{\pm}| \psi_{\pm} \rangle = 
 \langle \psi | 1 \pm  \gamma_5 | \psi \rangle \; ,
\label{eq:unique}
\end{equation}
where the scalar products are meant in colour and spin space, on an individual
lattice site. The analysis is carried out on 1 \% of the lattice sites with
maximal values $\langle\psi|\psi\rangle$.  As a result they produce histograms
of $X$-distributions from a number of near zero-eigenmodes and gauge
configurations.  This is motivated by a continuum consideration according to
which QCD instantons produce distributions in the quantity $X$ that peak at $X
= \pm 1$.

This procedure can be applied directly to normal Ginsparg-Wilson fer\-mions
where the choice of the bra-vector $\langle\psi|$ in Eq.~(\ref{eq:unique}) is
not to be questioned since the left and right eigenvectors coincide. The
situation is less clear once the normality is lost.  In our notation
Horv\'ath et al.~start
from right eigenmodes and use\footnote{I.~Horv\'ath,
  private communication}
\begin{equation}
  \tan  \left(\frac{\pi}{4}(1 + X) \right) =
  \frac{|\langle R_i|1-\gamma_5|R_i\rangle|^{1/2}}
  {|\langle R_i|1+\gamma_5|R_i\rangle|^{1/2}}\; .
\label{eq:horvath}
\end{equation}

For non-normal operators, however, the question arises whether 
it would be more economical  to adjust to the above described  biorthogonal
structure of operator representations. 
 This 
suggests to replace in  Eq.~(\ref{eq:unique})
\begin{equation}
  \label{eq:lr-def}
  \langle\psi_\pm|\psi_\pm\rangle \rightarrow \langle
  L_i|1\pm\gamma_5|R_i\rangle \; ,
\end{equation}
for Wilson-type fermions.

We emphasize that this modified prescription will coincide with the previous
one in the continuum limit, $a\rightarrow 0$, where the Dirac operator is
normal. Nevertheless one might hope that a lattice adapted definition of $X$
will suppress lattice effects. With this motivation we propose here to make
use of an ``improved'' quantity $X_{\mbox{\tiny imp}}$,
\begin{equation}
  \label{eq:ximp-def}
  \tan  \left(\frac{\pi}{4}(1 + X_{\mbox{\tiny imp}}) \right) =
  \frac{|\langle L_i|1-\gamma_5|R_i\rangle|^{1/2}}
  {|\langle L_i|1+\gamma_5|R_i\rangle|^{1/2}}\; .
\end{equation}
The selection of test points is consistently chosen 
with respect to 
$ |\langle L_i | R_i \rangle|$ (instead
of $\langle R_i | R_i \rangle$).

Throughout this letter, we shall refer to the two definitions as follows:
\begin{eqnarray}
\langle R_i | 1 \pm  \gamma_5 | R_i \rangle 
&\cong& \mbox{$RR$ definition},\\
\langle L_i | 1 \pm  \gamma_5 | R_i \rangle 
&\cong& \mbox{$LR$ definition}.
\end{eqnarray}

\section{Results for QED2}
\label{sec:results-qed2}
Let us consider two-dimensional quantum electrodynamics (QED2) in order
to probe the concept of the ``improved'' local chiral orientation parameter
$X_{\mbox{\tiny imp}}$.  In QED2 the gauge field is (anti)self-dual and one
should therefore observe peaks in the chiral orientation near plus and minus
one.
\begin{figure}[htb]
  \begin{center} \epsfig{file=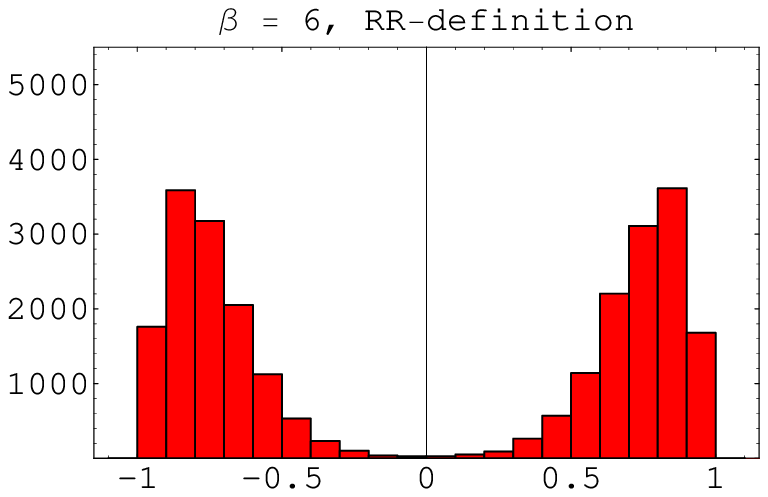, width=6 cm}
    \epsfig{file=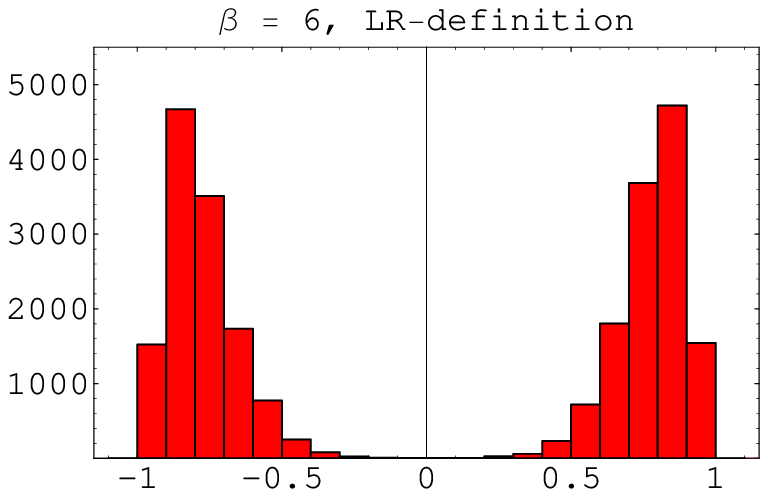, width=6 cm}
    \epsfig{file=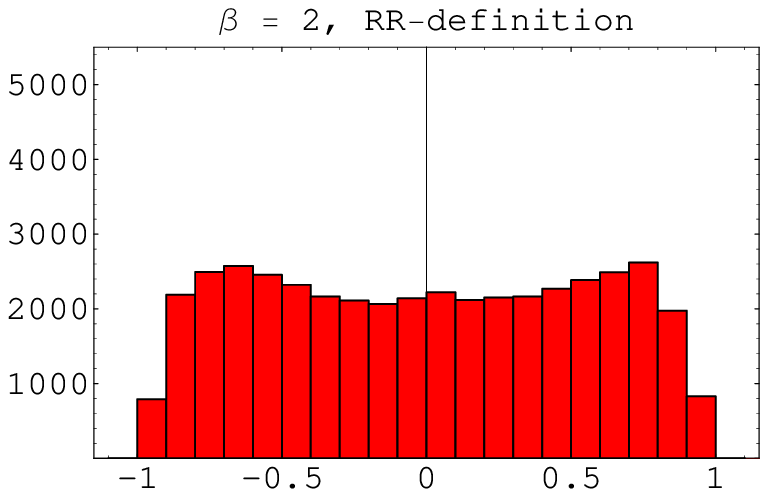, width=6 cm}
    \epsfig{file=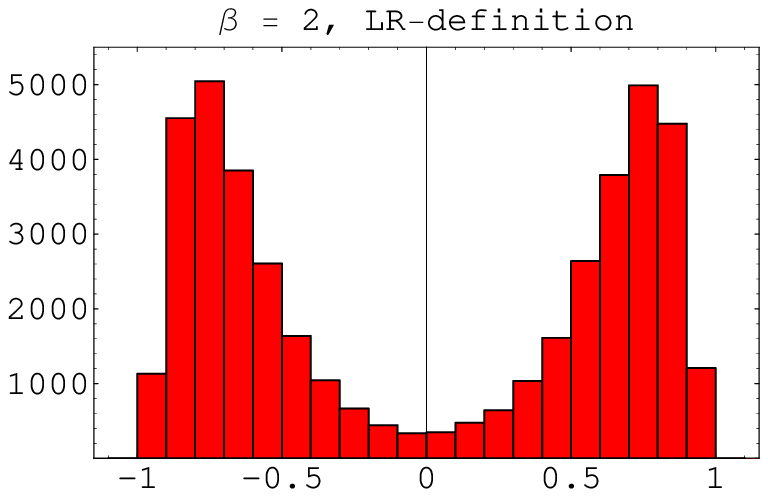, width=6 cm}
    \epsfig{file=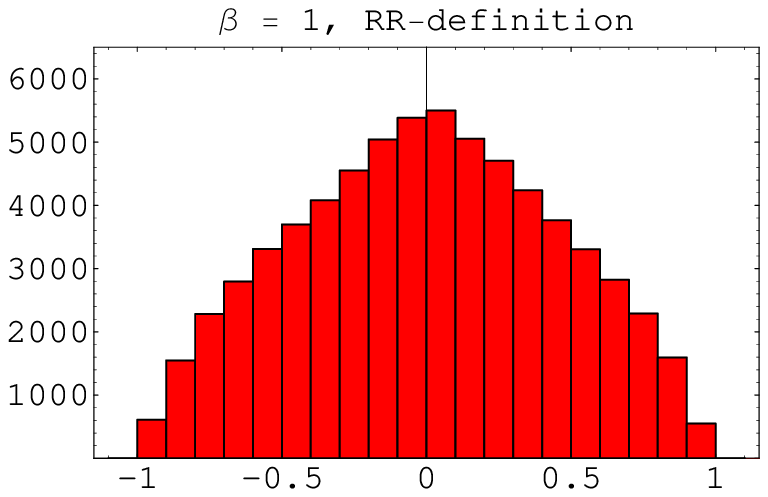, width=6 cm}
    \epsfig{file=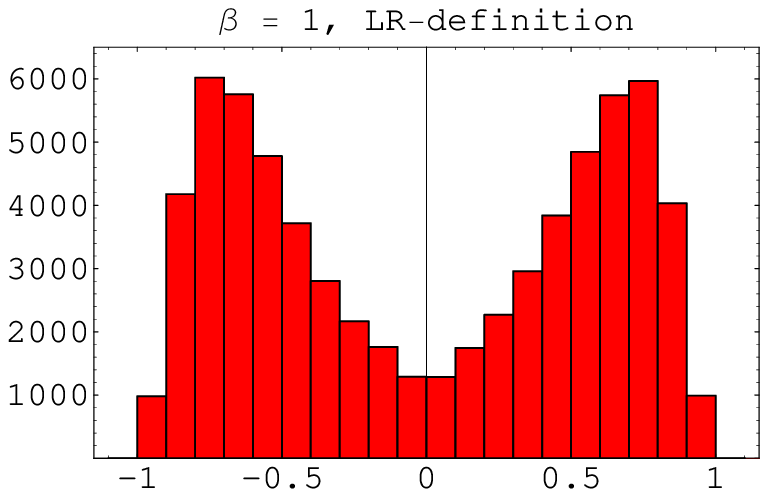, width=6 cm} \end{center}
    \caption
{\label{fig-sm-near-zero} Histograms of the local chiral
    orientation parameter for (complex) near-zero eigenvectors in QED2 for
    standard Wilson-Dirac fermions.  The two definitions, $RR$ and
    $LR$, have been used at $\beta = 6.0$, $2.0$ and $1.0$.}
\end{figure}     

In order to test the $LR$ definition, Eq.~(\ref{eq:ximp-def}), we perform an
analysis on a $24^2$ lattice for the $\beta$-values $6.0$, $2.0$ and $1.0$.
We determined the complex eigenmodes in the physical branch with
$\vert\mbox{Im}\,\lambda\vert<0.3$.  The resulting histograms for $500$ quenched
gauge field configurations for each $\beta$ are shown in the right column of
Fig.~\ref{fig-sm-near-zero}. Throughout our $\beta$-range we are able to
observe a very clear double peak structure.  For comparison, we also included
the histograms based on the $RR$ definition, Eq.~(\ref{eq:horvath}), for the
same sets of configurations. We find that the signal becomes almost invisible at
$\beta = 2.0$ and vanishes below. We emphasize that the results from both
definitions tend to coincide for small lattice spacings, as they should. On
the other hand, one can easily be misled by using the $RR$ definition at too
coarse lattice spacings.
\begin{figure}[htb]
  \begin{center} \epsfig{file=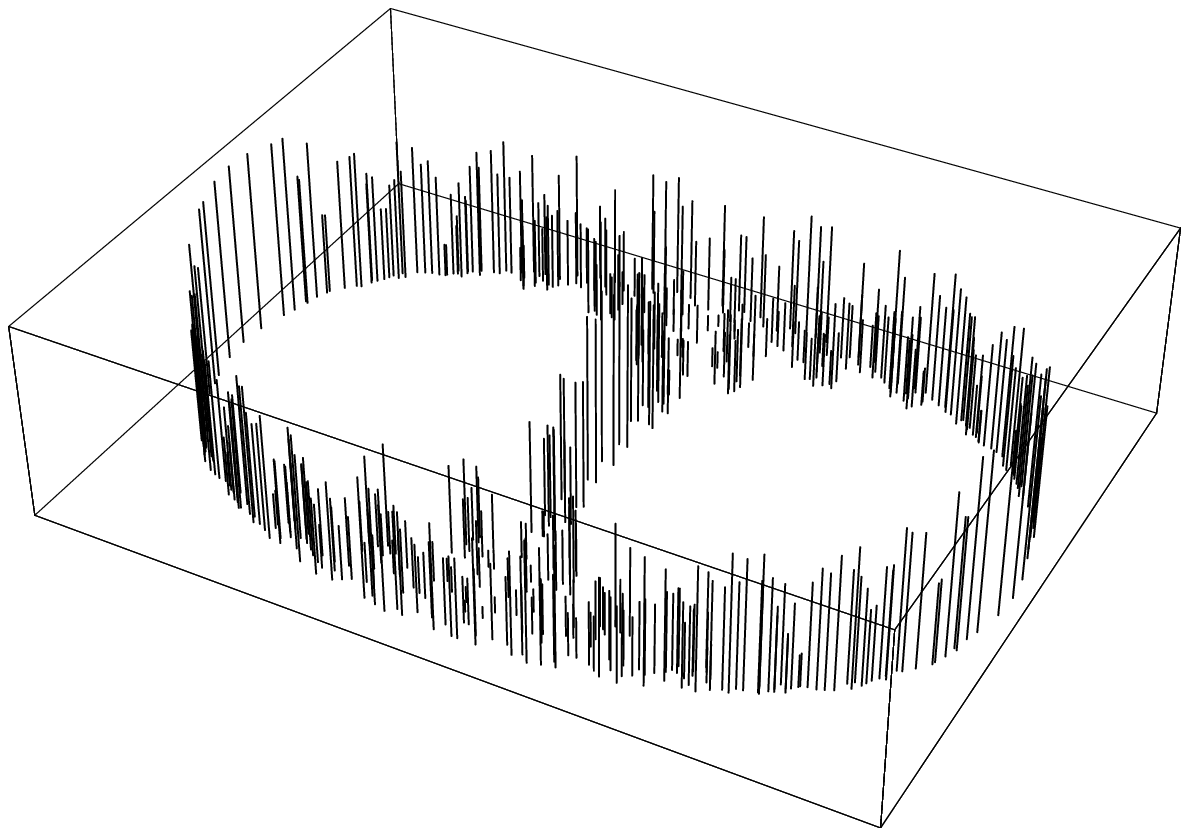, width=5.1 cm}
    \epsfig{file=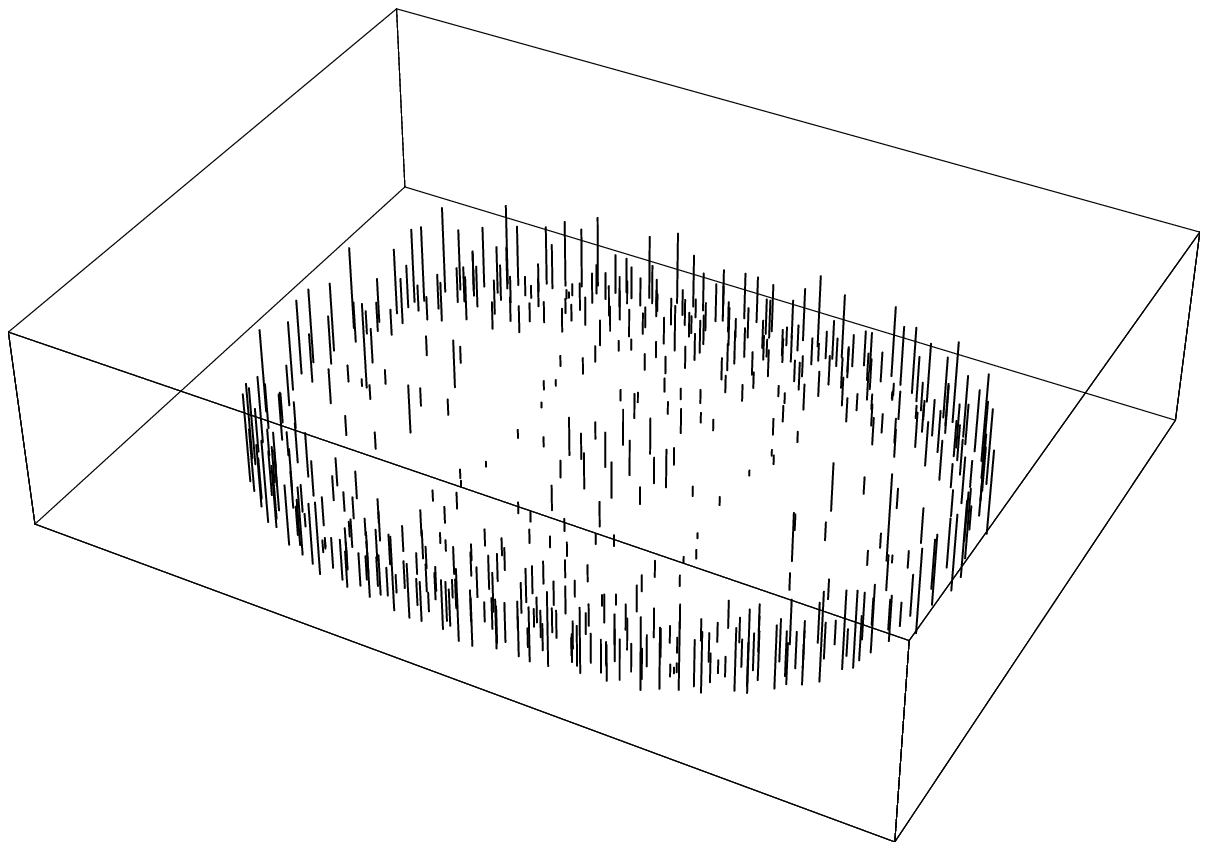, width=5.1 cm} \end{center}
    \caption{\label{fig:nep6vs1} Needle  plots of the quantity $\{ L_i
    , R_i \}$ (Eq.~(\ref{eq:angle})) computed for entire spectra on
    $16^2$ lattices at $\beta = 6.0$ (left) and $1.0$ (right).  The
    lengths of the vertical needles  represent the values of $\{ L_i ,
    R_i \}$ for eigenmode $i$, while their positions are given in the
    complex $\lambda_i$ plane. }
\end{figure}
\begin{figure}[!hb]
  \begin{center}
    \epsfig{file=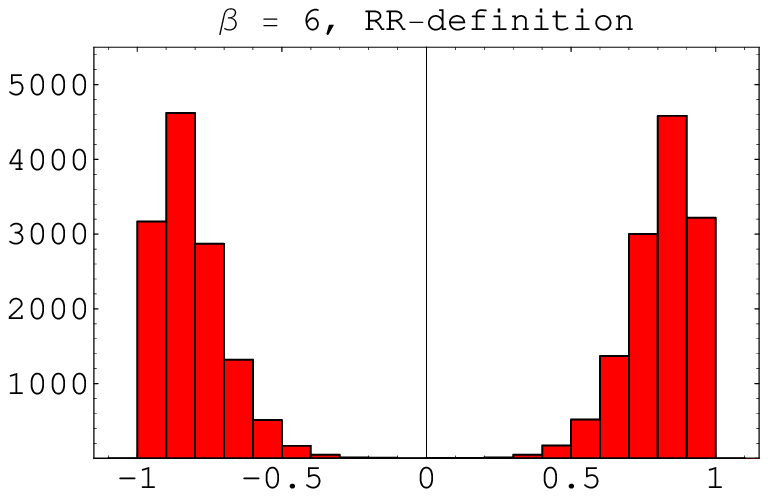, width=6 cm}
    \epsfig{file=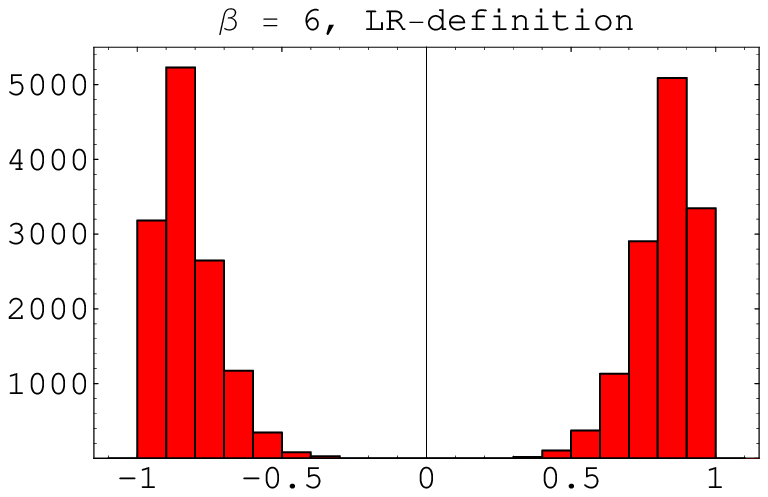, width=6 cm}
    \epsfig{file=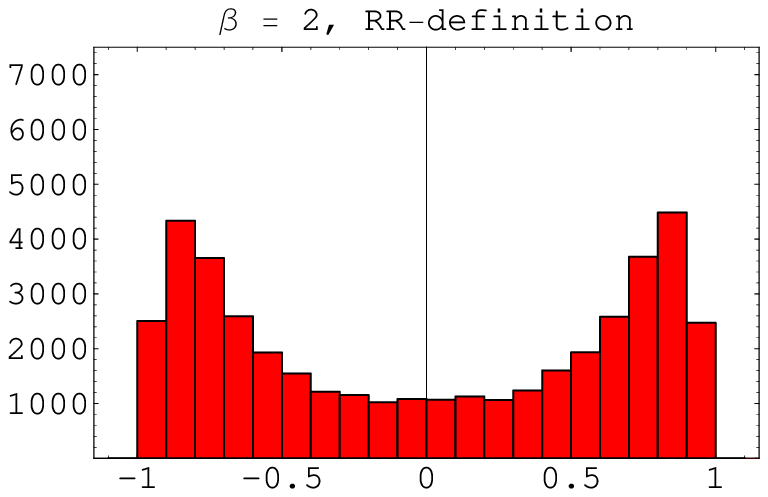, width=6 cm}
    \epsfig{file=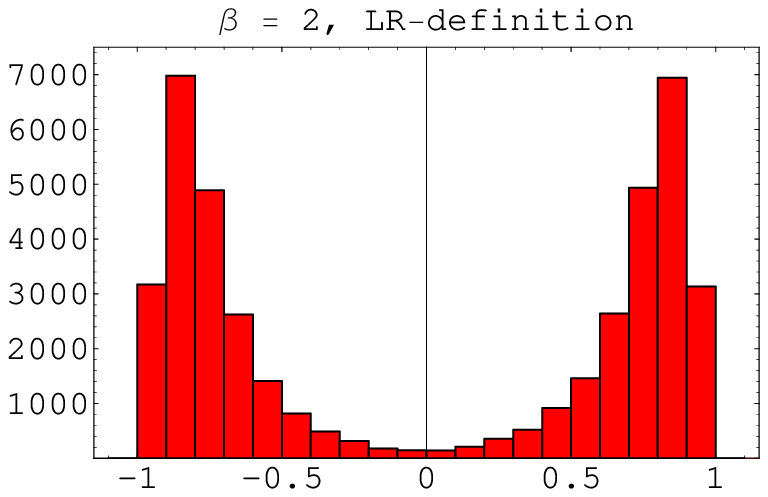, width=6 cm}
    \epsfig{file=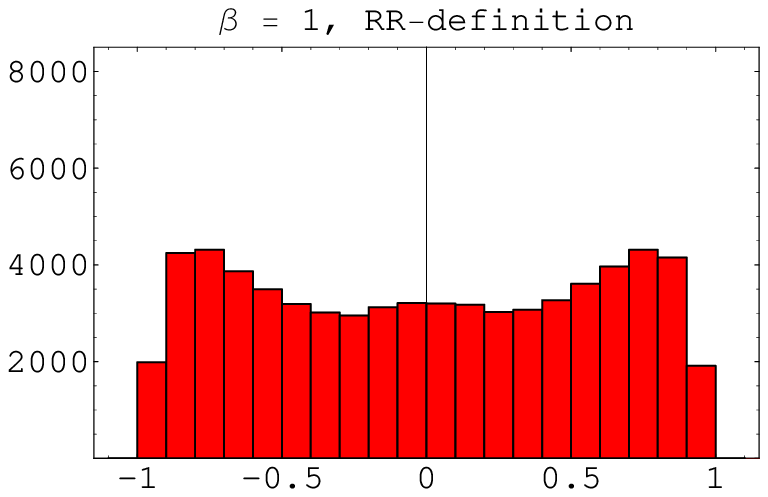, width=6 cm}
    \epsfig{file=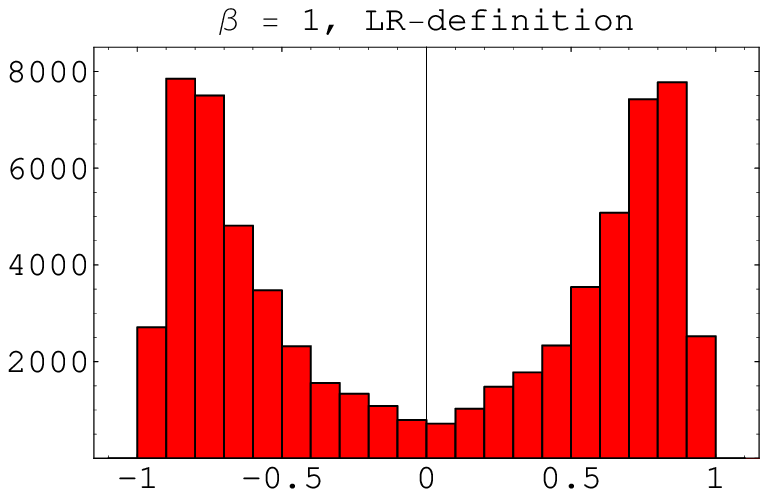, width=6 cm}
  \end{center}
  \caption{\label{fig:qed2-near-zero-clover} 
    Same as Fig.~\ref{fig-sm-near-zero} but with clover improved Wilson-Dirac
    fermions.}
\end{figure}

The reason behind this remarkable difference between the two schemes can be
traced to the loss of orthonormality of the right eigenvectors. For if the
left and hermitian conjugate right eigenmodes coincide, the following
``cosine''
\begin{equation}
\{ L_i , R_i \} \equiv \frac{\langle L_i | R_i \rangle} {\sqrt{\langle
   L_i | L_i \rangle} \sqrt{\langle R_i | R_i \rangle}}
   \label{eq:angle}
\end{equation}
would be equal to {\it one}. In Fig.~\ref{fig:nep6vs1} we show this quantity
in form of needle plots for the entire spectrum of eigenmodes for a $16^2$
lattice at $\beta = 6.0$ (left) and $1.0$ (right) in form of vertical lines
positioned at the complex eigenvalues.  For a normal operator,
$\{ L_i , R_i \}$ is identical to $1$ (for all $i$). Hence deviations
from {\it one} give us a measure for non-normality.

The values of $\{ L_i, R_i \}$ are found to be significantly closer to
{\it one} for the larger $\beta $-value. This holds in particular in
the physical branch (near the left edge of the box). In fact, at the lower
$\beta$-value, we are too far away from {\it one} as to still rely on the
orthogonality of the right eigenvectors alone.  

As another check of the consistency of this picture we have examined to
what extent an improved action helps to enhance the signal. To this end we
have computed the chiral orientation for the same set of quenched
configurations but with clover improved Wilson-Dirac fermions.  The results
are gathered in Fig.~\ref{fig:qed2-near-zero-clover} which is to be compared
directly to Fig.~\ref{fig-sm-near-zero}.  We find that the improvement on the
signal is fully consistent with the idea of accelerated convergence
$\langle R_i| \rightarrow \langle L_i|$.

\section{Results for QCD}
\label{sec:results-qcd}
\begin{figure}[!hb]
  \begin{center}
    \epsfig{file=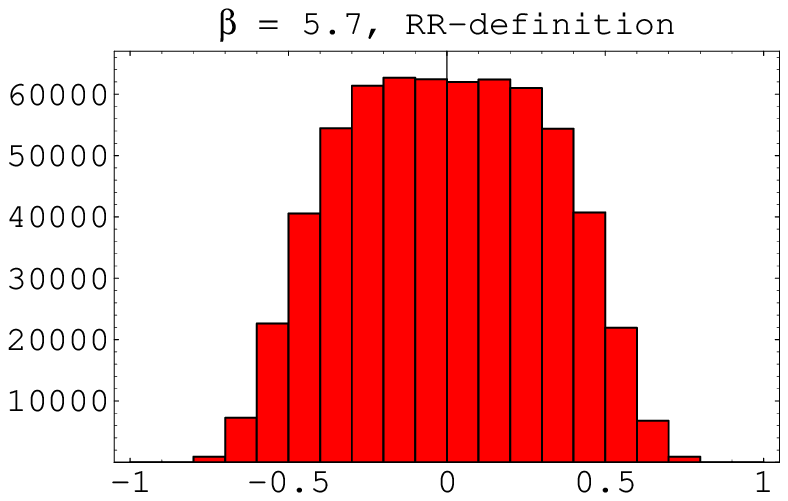, width=6 cm}
    \epsfig{file=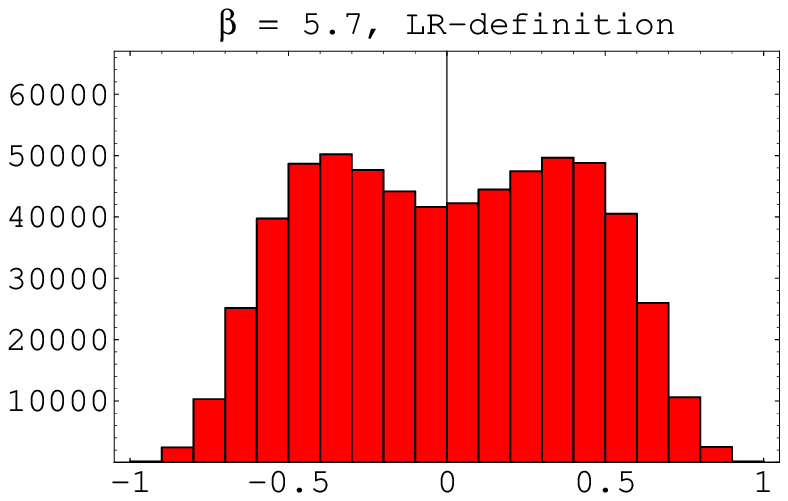, width=6 cm}
  \end{center}
  \caption{\label{fig-qcd57-near-real} The histograms for $X$ for the two
    definitions at $\beta=5.7$ for QCD for near-zero eigenvectors
    of the Wilson-Dirac matrix, $D_W$.}
\end{figure}

\begin{figure}[htb]
  \begin{center}
    \epsfig{file=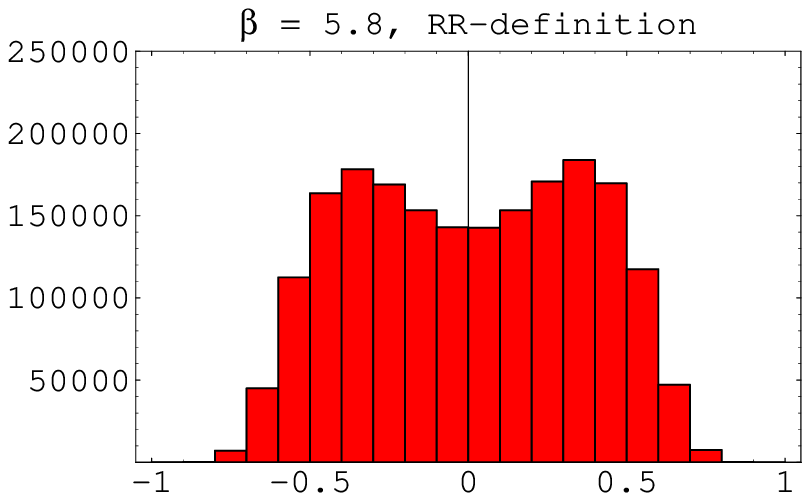, width=6 cm}
    \epsfig{file=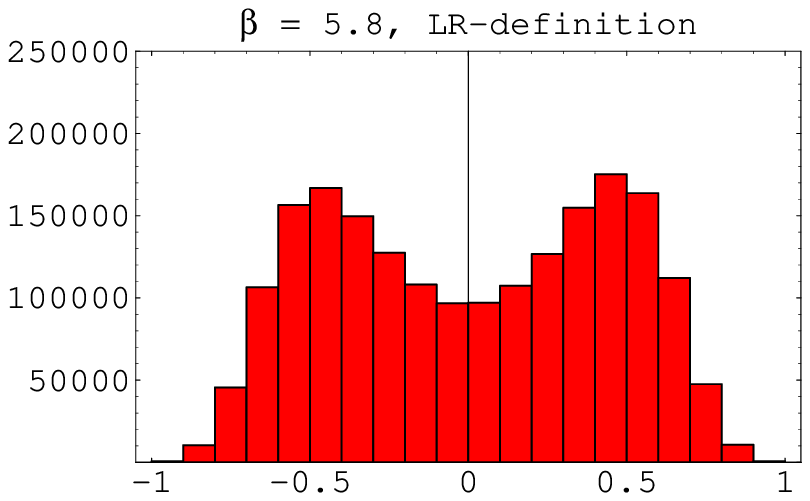, width=6 cm}
    \epsfig{file=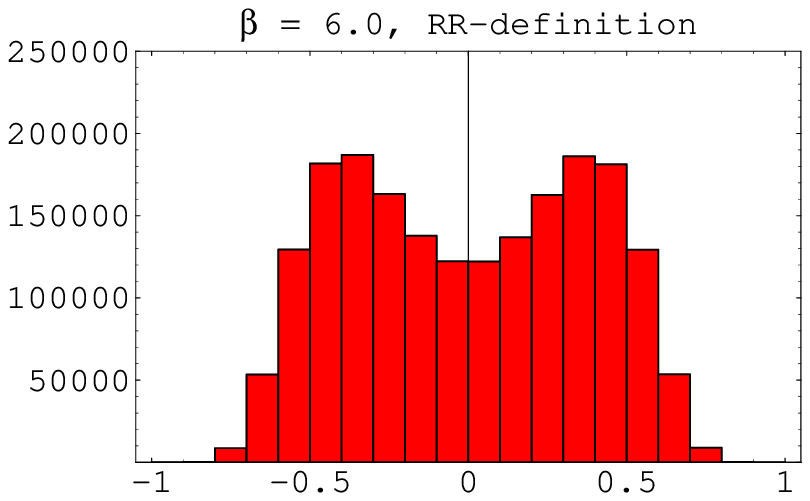, width=6 cm}
    \epsfig{file=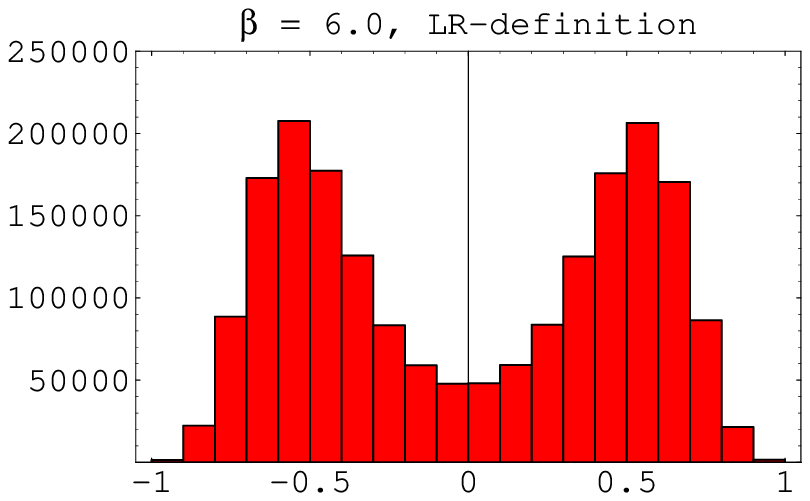, width=6 cm}
    \epsfig{file=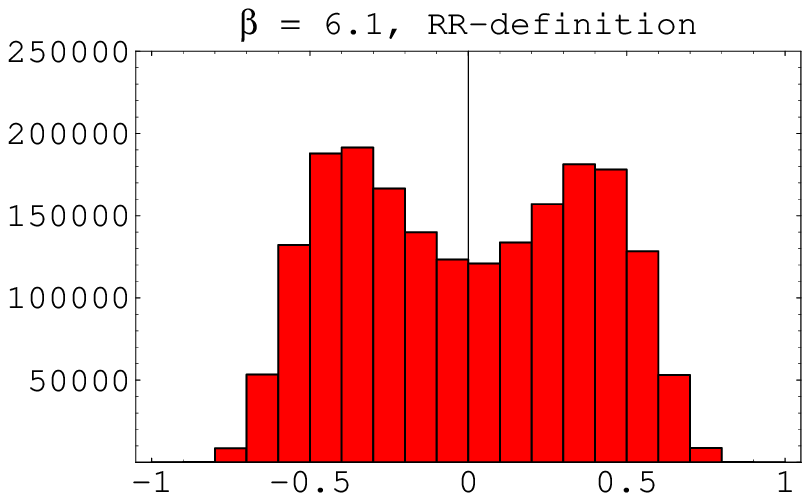, width=6 cm}
    \epsfig{file=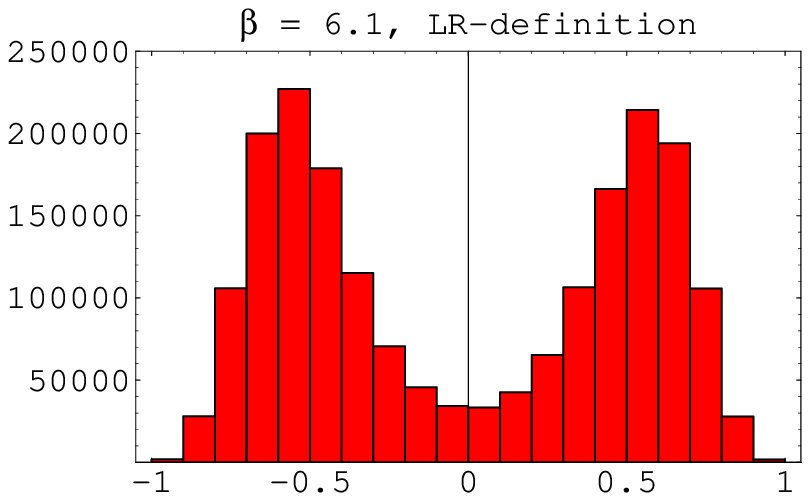, width=6 cm}
  \end{center}
  \caption{\label{fig-qcd-near-real} The histograms for $X$ for the two
    definitions at $\beta=5.8$, $\beta=6.0$ and $\beta=6.1$ for
    QCD for near-zero eigenvectors of the Wilson-Dirac matrix, $D_W$.}
\end{figure}

Encouraged by our results in QED2 we shall now turn to quenched
QCD. To compare with the results of Ref.~\cite{Horvath:2001ir}, we
computed 50 near-zero (complex) eigenmodes on 30 configurations of
size $12^3\times 24$ at the same $\beta$-value, $\beta = 5.7$. The
resulting chirality distributions are plotted in
Fig.~\ref{fig-qcd57-near-real}, for both the $RR$ and the $LR$ definitions.
While the $RR$-result shows only a plateau very similar to the one
seen  in Ref.~\cite{Horvath:2001ir}, we do find a slight indication
of a double peak structure in the $LR$ case.
As a next step, we proceed to finer lattices, corresponding to $\beta = 5.8$,
$6.0$ and $6.1$, on $16^3\times 32$ configurations. The results from the $RR$
and $LR$ definitions are confronted in Fig.~\ref{fig-qcd-near-real}.  We see
unambiguous evidence for a deepening of the valley between two clear peaks,
which appear to move outwards, towards $\pm 1$.  This is confirmed by the
results from the $RR$ definition, though with quite some delay when increasing
$\beta$. In order to achieve the same signal quality one has to work with
$\beta = 6.1$ ($RR$), instead of $\beta = 5.8$ ($LR$).

\begin{figure}[htb]
  \begin{center}
    \epsfig{file=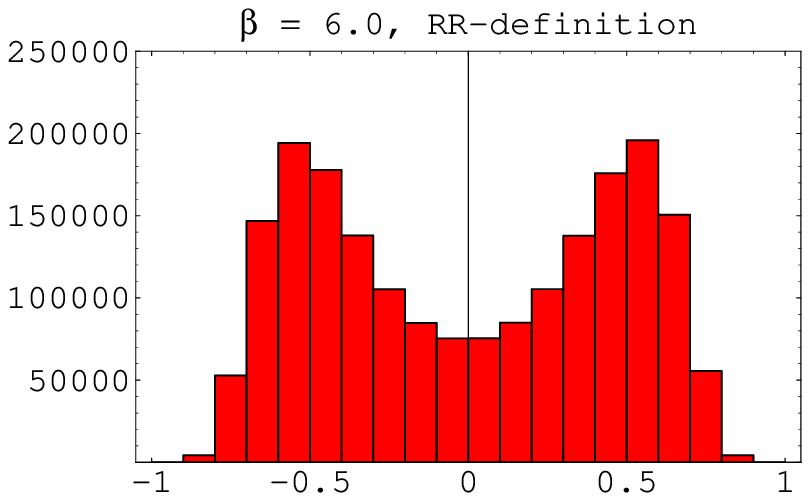, width=6 cm}
    \epsfig{file=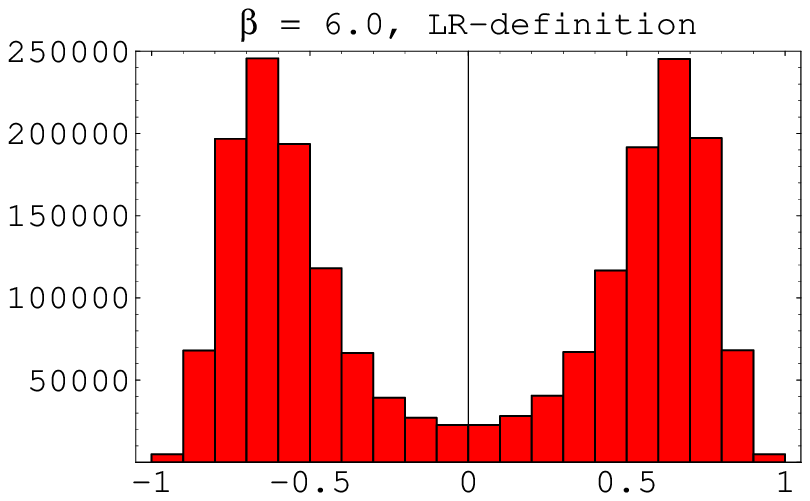, width=6 cm}
  \end{center}
  \caption{\label{fig-qcd-near-real-clover} The histograms for $X$ at
    $\beta=6.0$ in QCD for the two definitions using the near-zero
    eigenvectors of the clover-improved Wilson-Dirac operator,
    $D_{CW}$.}
\end{figure}

To complete the study we repeated the computation with clover
improvement, at $\beta = 6.0$. The histograms are shown in
Fig.~\ref{fig-qcd-near-real-clover}. As in the QED2 case we observe a
markedly   better  separation of the peaks in the $RR$ situation.
There is also some further signal improvement in the $LR$ case.

\section{Summary and conclusion}
In their debate about the conclusiveness of the local chirality data
presented in Ref.~\cite{Horvath:2001ir}, the authors
of~\cite{DeGrand:2001pj} have suspected that the artifacts and the use of
a non-chiral lattice action might hide  the signal for correlations
between instanton positions and wave function localizations. This is
precisely what we confirm in our analysis.

To clarify the situation, we studied finer lattices and repaired for
the loss of orthogonality due to the non-normality of the Wilson-Dirac
matrix. As a result we do see a clear double peak structure  on
lattices with resolutions higher than 0.1 fm ($\beta \geq 6.0$).  We
found that the lattice artifacts can be considerably reduced by
exploiting the biorthogonal system of left and right eigenvectors.

We conclude that the dominance of instantons on topological charge fluctuations
is not at all ruled out by local chirality measurements. 

In order to consolidate this picture it appears worthwhile to
investigate which fermion formulation is actually the most appropriate in this
context: although the overlap fermions seem to yield convincing peaks
it is not clear whether less local actions might wipe out
the fluctuations that one is after.
 But now the door is
open for investigations of the local chiral structure, relying on
ultralocal probes. This will help to elucidate the significance of the
overlap signal.

\vskip5mm
\noindent
\textbf{Acknowledgements: } The authors would like to thank Andreas
Frommer, Christof Gattringer, Ivan Horv\'ath and Tamas Kovacs for
useful discussions and comments.  We thank the EU network ``Hadron
phenomenology from Lattice QCD'' (HPRN-CT-2000-00145) for providing the
stimulating atmosphere for these discussions at the EU Joint Training
Course ``Algorithms, Actions and Computers''. WS appreciates support
from the DFG Graduiertenkolleg ``Feldtheoretische und Numerische
Methoden in der Statistischen und Elementarteilchenphysik''. The
sets of quenched gauge field configurations for $\beta = 5.8$ and
$\beta = 6.0$ were downloaded from the "Gauge Connection" WEB-site at
NERSC (\texttt{http://qcd.nersc.gov/}).  We thank the DFG (grant Li
701/3-1) for supporting the parallel computer ALiCE. The 
computations were done on ALiCE in Wuppertal and CRAY T3E system of ZAM at
Research Center J\"ulich.

\end{document}